\pdfoutput=1
\documentclass[preprint]{elsarticle}
\usepackage{hyperref}
\usepackage{amsmath}
\usepackage{bm}
\usepackage{subcaption}

\def\mydot{\!\cdot\!}

\usepackage{color}

\begin{document}

\begin{frontmatter}

\newcommand{\reportnumber}[1]{%
  \makebox[0pt][l]{%
    \hspace{1cm}%
    \raisebox{3em}[0pt][0pt]{%
      \parbox[t][0pt][b]{\textwidth}{%
        \begin{flushright}%
          \normalsize #1
        \end{flushright}%
      }%
    }%
  }%
  \hskip 0pt plus 0.0001fil
}

\title{\reportnumber{NIKHEF-2015-017}
	The diamond rule for multi-loop Feynman diagrams}

\author[n,l]{B.~Ruijl}
\ead{benrl@nikhef.nl}

\author[n]{T.~Ueda}
\ead{tueda@nikhef.nl}

\author[n]{J.A.M.~Vermaseren}
\ead{t68@nikhef.nl}

\address[n]{Nikhef Theory Group,
         Science Park 105, 1098 XG Amsterdam, The Netherlands}
\address[l]{Leiden University, Niels Bohrweg 1, 2333 CA Leiden, The Netherlands}

\begin{abstract}
An important aspect of improving perturbative predictions in high energy 
physics is efficiently reducing dimensionally regularised Feynman integrals 
through integration by parts (IBP) relations. The well-known triangle rule 
has been used to achieve simple reduction schemes. In this work we 
introduce an extensible, multi-loop version of the triangle rule, which we 
refer to as the diamond rule. Such a structure appears frequently in 
higher-loop calculations. We derive an explicit solution for the recursion, 
which prevents spurious poles in intermediate steps of the computations. 
Applications for massless propagator type diagrams at three, four, and 
five loops are discussed.
\end{abstract}

\begin{keyword}
  Feynman integrals \sep integration by parts identities
\end{keyword}

\end{frontmatter}

\section{Introduction}
\begin{figure}
\centering
\includegraphics[scale=0.8]{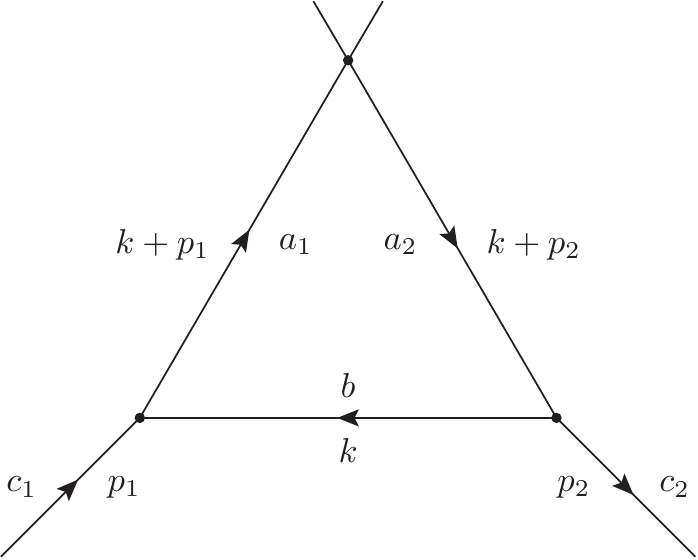}
\caption{\label{fig:triangle}A triangle subtopology where the loop momentum $k$ 
  is assigned to the central line. 
  $p_1$ and $p_2$ are external momenta.
  $a_1$, $a_2$, $b$, $c_1$, and $c_2$ represent the powers of their associated 
  propagators.}
\end{figure}

Reducing complexities of Feynman integrals through integration by parts 
(IBP) relations~\cite{Chetyrkin:1981qh,Tkachov:1981wb} is an important 
component of modern multi-loop calculations. Finding more efficient 
reduction methods allows the computation of higher order terms in 
perturbative expansions which in turn aids in providing a better 
quantitative understanding of ongoing experiments. Since the 1980s, the 
so-called \emph{triangle rule}~\cite{Chetyrkin:1981qh,Tkachov:1981wb} has 
been used for removing a propagator line from diagrams corresponding to a 
certain class of integrals. Any topology that has the following 
substructure can be simplified using the triangle rule:
\begin{align}
\begin{split}
  F(&a_1,a_2,b,c_1,c_2)
  =\\
  &\int d^D k
  \frac{
    k^{\mu_1} \dots
    k^{\mu_N}
  }{
    \bigl[(k+p_1)^2 + m_1^2 \bigr]^{a_1}
    \bigl[(k+p_2)^2 + m_2^2 \bigr]^{a_2}
    (k^2)^{b}
    (p_1^2 + m_1^2)^{c_1}
    (p_2^2 + m_2^2)^{c_2}
  }
  ,
\end{split}
  \label{eq:triangle-int}
\end{align}
where $D$ is the dimension which is set to $4-2\epsilon$~\cite{Bollini:1972ui,tHooft:1972fi}, and 
$b$, $c_1$, $c_2$ are positive integers. The diagram corresponding to this 
integral is shown in Fig.~\ref{fig:triangle}.

We write out the IBP relation $\frac{\partial}{\partial k_\mu} k_\mu F=0$ 
(where the derivative must be performed before the integration) to obtain
\begin{equation}
  1
  =
  \frac{1}{D + N - a_1 - a_2 - 2 b}
  \Bigl[
    a_1 \bm{A}_1^+ ( \bm{B}^- - \bm{C}_1^- )
    + a_2 \bm{A}_2^+ ( \bm{B}^- - \bm{C}_2^- )
  \Bigr]
  ,
  \label{eq:triangle-id}
\end{equation}
where $\bm{A}_i^+$, $\bm{B}^-$, and $\bm{C}_i^-$ are operators acting on an 
integral that increase the power $a_i$ by one, decrease the power $b$ by 
one, and decrease the power $c_i$ by one, respectively. Numerators that are 
expressed in dot products of $k$ and an external line, contribute as a constant $N$ to the rule.
The rule of the triangle can be recursively applied to remove one of the 
propagators associated with $k$, $p_1$, or $p_2$ from the system.

The recursion in the triangle rule can be explicitly 
`solved'~\cite{Tkachov:1984xk}, such that the solution is expressed as a 
linear combination of integrals for which either $b$, $c_1$, or $c_2$ is 0. 
The advantage of the summed system over the recursion is that 
it generates fewer intermediate terms and it cannot 
have \emph{spurious poles}: terms in which the factor $D + N - a_1 - a_2 - 
2 b$ becomes proportional to $\epsilon$ more than once during the full 
recursion.

In this work we introduce a more general class of diagrams that can be 
reduced using an extension of the triangle rule. We call this rule the 
\emph{diamond rule}. We will show that the diamond rule 1) can be extended 
to any number of loops, 2) allows for a complete set of irreducible numerators that 
only contributes as a constant  
and 3) can be explicitly `solved' to prevent spurious poles.

It should be noted that throughout this paper we mean by reduction the 
removal of a single line. This does not necessarily mean that the remaining 
diagram(s) will be trivial. An example of this is the four-loop ladder 
diagram, which can be reduced twice with the rule of the triangle, after 
which one of the remaining diagrams involves a master integral and needs a 
complete reduction scheme of 14 steps in which all numerators are removed 
and the power of all denominators is lowered to one successively.

In Section~\ref{sec:diamond} the diamond rule is derived. 
Section~\ref{sec:summation} shows the explicit summation formula and 
Section~\ref{sec:examples} shows examples. Finally, 
Section~\ref{sec:conclusion} gives a conclusion and discussion.

\section{Diamond rule}
\label{sec:diamond}

\begin{figure}
  \centering
  \includegraphics{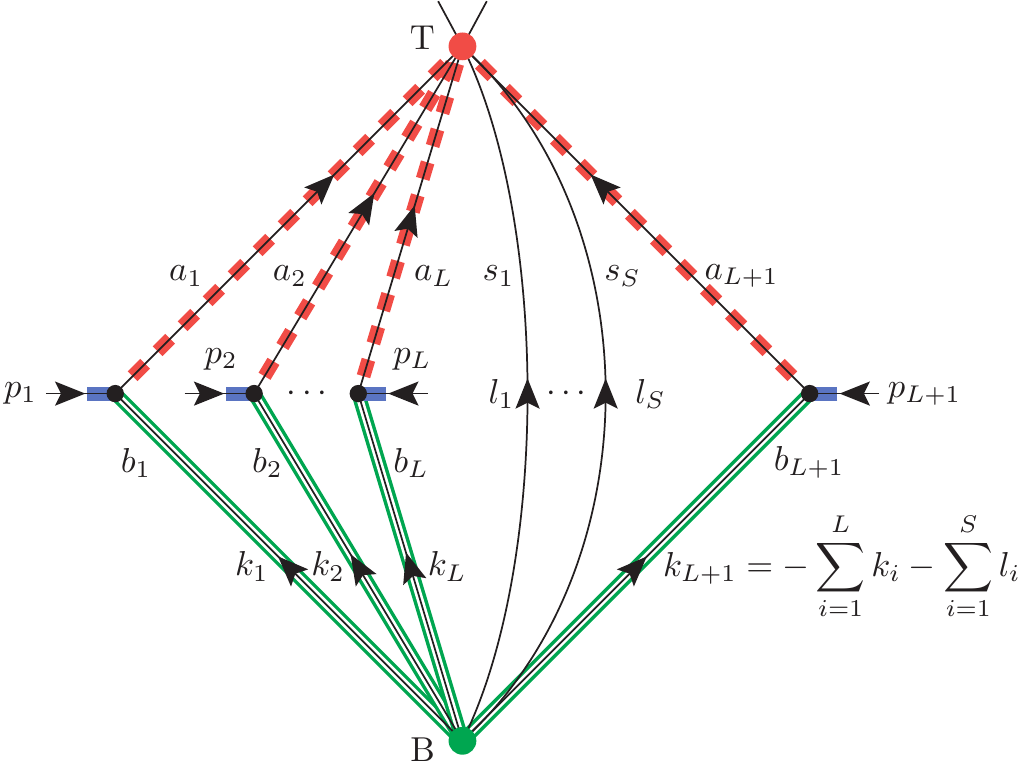}
  \caption{$(L+S)$-loop diamond-shaped diagram. $(L+1)$-lines have external
    connections and $S$-lines do not.
    Red with dashed lines, green with double lines, and blue with 
thick lines represent upper, lower, and external lines of the diamond, 
respectively. Label $T$ represents the top vertex, and $B$ the bottom 
vertex. $k_i$, $p_i$, and $l_i$ are momenta, and $a_i$, $b_i$, and $s_i$ are 
the powers of their associated propagators.
  }
  \label{fig:ldiamond}
\end{figure}

Consider the following family of Feynman integrals in $D$-dimensions 
arising from the $(L+S)$-loop diagram in Fig.~\ref{fig:ldiamond}:
\begin{multline}
  F\bigl(\{a_i\},\{b_i\} \bigr)
  =
  \left[
  \prod_{i=1}^{L}
  \int d^D k_i
  \right]
  \left[
  \prod_{i=1}^{S}
  \int d^D l_i
  \right]
  \\
  \times
  \left[
  \prod_{i=1}^{L+1}
  \frac{
    k_i^{\mu^{(i)}_1} \dots
    k_i^{\mu^{(i)}_{N_i}}
  }{
    \bigl[(k_i+p_i)^2 + m_i^2 \bigr]^{a_i}
    (k_i^2)^{b_i}
  }
  \right]
  \left[
  \prod_{i=1}^S
  \frac{
    l_i^{\nu^{(i)}_1} \dots
    l_i^{\nu^{(i)}_{R_i}}
  }{
    (l_i^2)^{s_i}
  }
  \right]
  .
  \label{eq:diamond-int}
\end{multline}

The diagram consists of $(L+1)$ paths from the top vertex $T$ to the bottom 
vertex $B$ with an external connection in between, and $S$ lines without 
external connections. The upper, lower, and external lines of the diamond are represented by 
red with dashed lines, green with double lines, and blue with thick lines, respectively. 
The lines without external connections, we call \emph{spectator lines}. In 
principle any pair of spectator lines can be seen as a two point function 
which can be reduced to a single line by integration. This line would then 
have a power that is not an integer. Depending on the complete framework of 
the reductions this may or may not be desirable. Hence we leave the number 
of spectators arbitrary. In any case, the contribution of 
the spectators is a constant (see below), which allows us to characterise 
integrals in the family only by $2(L+1)$ indices $a_i$ and $b_i$, and not 
by $s_i$. Without loss of generality, we assign loop momenta $k_i$ to the 
lower lines of the diamond as well as $l_i$ to the spectator lines, except 
the last diamond line which is fixed by momentum conservation:
\begin{equation}
  k_{L+1} = - \sum_{i=1}^L k_i - \sum_{i=1}^S l_i .
\end{equation}
In contrast, we do not require any constraints on the momentum 
conservation at the top vertex in the arguments below, hence any number of 
external lines can be attached to this point. In the middle of the diamond, 
external lines with momentum $p_i$ are attached by three-point vertices. 
The upper lines in the diamond may have masses $m_i$, whereas the lower 
lines in the diamond and the spectator lines have to be massless. In 
addition, we allow arbitrary tensor structures of $k_i$ and $l_i$ with 
homogeneous degrees $N_i$ and $R_i$, respectively, in the numerator.

Constructing the IBP identity corresponding to the operator
\begin{equation}
    \sum_{i=1}^L \frac{\partial}{\partial k_i} \mydot k_i
  + \sum_{i=1}^S \frac{\partial}{\partial l_i} \mydot l_i
  ,
\end{equation}
straightforwardly gives the following operator identity:
\begin{equation}
  (L+S) D
  + \sum_{i=1}^{L+1} (N_i - a_i - 2 b_i)
  + \sum_{i=1}^S (R_i - 2 s_i)
  =
  \sum_{i=1}^{L+1} a_i \bm{A}_i^+ \bigl[ \bm{B}_i^- - (p_i^2 + m_i^2) \bigr]
  .
  \label{eq:diamond-id}
\end{equation}
Here $\bm{A}_i^+$ and $\bm{B}_i^-$ are understood as operators increasing 
$a_i$ and decreasing $b_i$ by one, respectively, when acting on $F(\{a_i\},
\{b_i\})$. Note that operators changing the spectator indices $s_i$ are 
absent in the identity.

For a typical usage of Eq.~\eqref{eq:diamond-id}, one may identify a 
diamond structure as a subgraph in a larger graph. If the line with the 
momentum $p_i$ has the same mass $m_i$ as the corresponding upper line, the 
term $(p_i^2+m_i^2)$ in the identity reads as an operator $\bm{C}_i^-$ 
decreasing the corresponding index $c_i$ of the power of the propagator 
$(p_i^2+m_i^2)^{-c_i}$ in the larger graph by one. Applying the rule
\begin{equation}
  1
  =
  \frac{1}{E}
  \sum_{i=1}^{L+1} a_i \bm{A}_i^+ ( \bm{B}_i^- - \bm{C}_i^- )
  ,
  \label{eq:diamond-rule}
\end{equation}
where
\begin{equation}
  E = (L+S)D + \sum_{i=1}^{L+1} (N_i - a_i - 2 b_i)
           + \sum_{i=1}^S (R_i - 2 s_i) ,
\end{equation}
decreases $\sum_{i=1}^{L+1} (b_i + c_i)$ of integrals appearing in the 
right-hand side, at the cost of increasing $\sum_{i=1}^{L+1} a_i$. Starting 
from positive integer indices $b_i$ and $c_i$, one can repeatedly use the 
rule until one of either $b_i$ or $c_i$ is reduced to zero.%
\footnote{
  Note that $a_i$ are allowed to be non-integers provided the denominator
  in the right-hand side
  of Eq.~\eqref{eq:diamond-rule} never vanishes.
}%

The above diamond rule contains the conventional
triangle rule as a special case.
For the one-loop case $L=1$ and $S=0$, the two lower lines may be 
identified as a single line and the triangle integral in Eq.~\eqref{eq:triangle-int} can 
be reproduced. Correspondingly, the IBP identity~\eqref{eq:diamond-rule} 
becomes Eq.~\eqref{eq:triangle-id}.

\section{Summation rule}
\label{sec:summation}
We now derive an explicit summation formula for the recursion in the 
diamond rule. First, we consider the possible connectivities. If we allow 
for some external lines to be directly connected to each other, we get at least one 
triangle that can be used for the triangle rule: suppose the 
external momenta of $k_i$ and of $k_j$ are connected and identified with 
$p_{ij}$, then this triangle is $k_i$, $k_j$, $p_{ij}$. In this case, the 
triangle rule generates fewer terms and is preferred to the diamond rule. 
Thus, we only consider the case where the diamond does not have direct connections of external lines.

We follow the same procedure as outlined in \cite{Tkachov:1984xk}. First, 
we rewrite Eq.~\eqref{eq:diamond-rule} as:
\begin{equation}
  F
  =
  \Biggl[\sum_{i=1}^{L+1} a_i \bm{A}_i^+ ( \bm{B}_i^- - \bm{C}_i^- )
  \Biggr] \bm{E}^{-1} F
  ,
  \label{eq:diamond-op}
\end{equation}
where $\bm{E}$ is the operator $(L+S)D + \sum_{i=1}^{L+1} 
(N_i - a_i - 2 b_i) + \sum_{i=1}^S (R_i - 2 s_i)$. We split our solution 
in two classes $\bm{A}_i^+ \bm{B}_i^-$ and $\bm{A}_i^+ \bm{C}_i^-$ satisfying
\begin{equation}
  \bm{E}^{-1}(\bm{A}_i^+ \bm{B}_i^-)=(\bm{A}_i^+
 \bm{B}_i^-)(\bm{E} + 1)^{-1},\qquad 
\bm{E}^{-1}(\bm{A}_i^+ \bm{C}_i^-)=(\bm{A}_i^+ 
\bm{C}_i^-)(\bm{E} - 1)^{-1} .
\end{equation}
 We identify the first class with the label $+$, since it increases 
$E$ by 1, and the latter with the label $-$, since it decreases 
$E$ by 1. The remaining part of the derivation is analogous to 
the one in \cite{Tkachov:1984xk}.

Finally, we obtain the explicit summation formula:
\begin{align}
\begin{split}
F(&\{a_i\}, \{b_i\},\{c_i\}) = \\
&\sum_{r=1}^{L+1} \Biggl[ \Biggl(\prod_{\substack{i=1\\i\neq r}}^{L+1} 
\sum_{k_i^+=0}^{b_i-1}\Biggr) \Biggl(\prod_{i=1}^{L+1} 
\sum_{k_i^-=0}^{c_i-1}\Biggr) 
(-1)^{k^-} \frac{k_r^+(k^+ + k^- -1)!}{\prod_{i=1}^{L+1} k_i^+!k_i^-!}
 (E+k^+)_{-k^+-k^-}\\
&\times \left( \prod_{i=1}^{L+1}(a_i)_{k_i^++k_i^-} \right) F
 \left(\{a_i+k_i^+ +k_i^-\}, \{b_i-k_i^+\},\{c_i-k_i^-\}\right)
 \Biggr]_{k_r^+=b_r}\\
+&\sum_{r=1}^{L+1} \Biggl[ \Biggl(\prod_{i=1}^{L+1} 
\sum_{k_i^+=0}^{b_i-1}\Biggr) \Biggl(\prod_{\substack{i=1\\i\neq r}}^{L+1} 
\sum_{k_i^-=0}^{c_i-1}\Biggr) 
(-1)^{k^-} \frac{k_r^-(k^+ + k^- -1)!}{\prod_{i=1}^{L+1} k_i^+!k_i^-!}
 (E+k^+ +1)_{-k^+-k^-} \\
&\times \left(\prod_{i=1}^{L+1} (a_i)_{k_i^++k_i^-} \right) F
 \left(\{a_i+k_i^+ +k_i^-\}, \{b_i-k_i^+\},\{c_i-k_i^-\}\right)
 \Biggr]_{k_r^-=c_r},\\
\end{split}
\end{align}
where $k^+=\sum_{i=1}^L k_i^+$, $k^-=\sum_{i=1}^L k_i^-$, and $(a)_b$ is 
the rising Pochhammer symbol $\Gamma(a+b)/\Gamma(a)$. The first term decreases 
the power $b_r$ to 0, and the second term decreases $c_r$ to 0. 
The only significant difference between the two terms is the $+1$ in the Pochhammer symbol.

Because the Pochhammer symbol that depends on $E$ only appears once in 
each term, powers of $1/\epsilon^2$ or higher cannot occur. Thus, the 
explicit summation formula for the diamond rule does not have spurious poles.

\section{Examples}
\label{sec:examples}
\begin{figure}
\centering
\begin{subfigure}{0.48\textwidth}
\includegraphics[width=\linewidth]{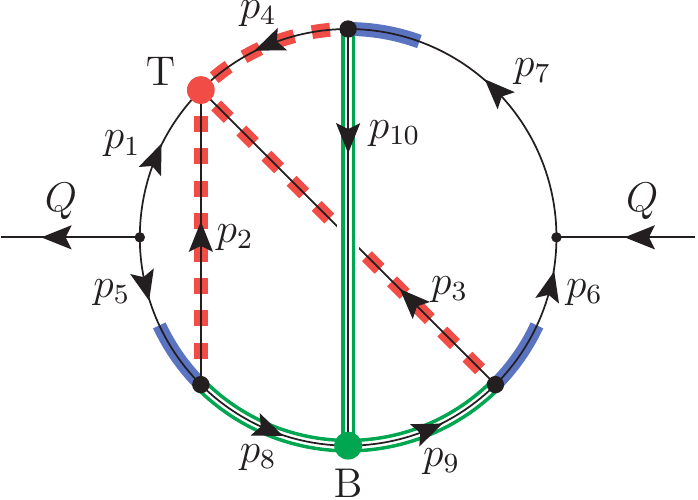}
\caption{\label{fig:no5}Four-loop topology}
\end{subfigure}
\quad
\begin{subfigure}{0.48\textwidth}{
\raisebox{6pt}{%
\includegraphics[width=\linewidth]{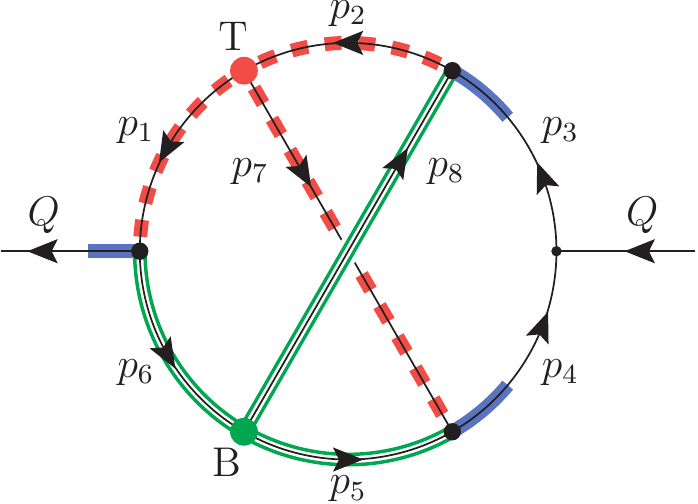}}}
\caption{\label{fig:no}Three-loop NO topology}
\end{subfigure}
\begin{subfigure}{0.96\textwidth}{
\includegraphics[width=0.48\linewidth]{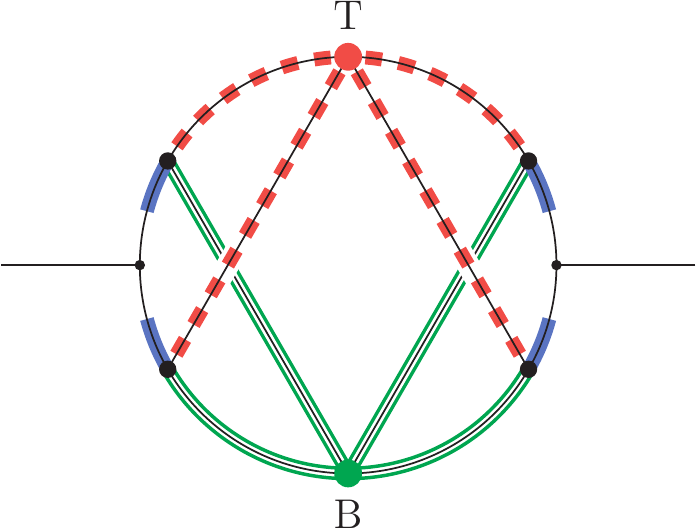}
\quad
\raisebox{12pt}{%
\includegraphics[width=0.48\linewidth]{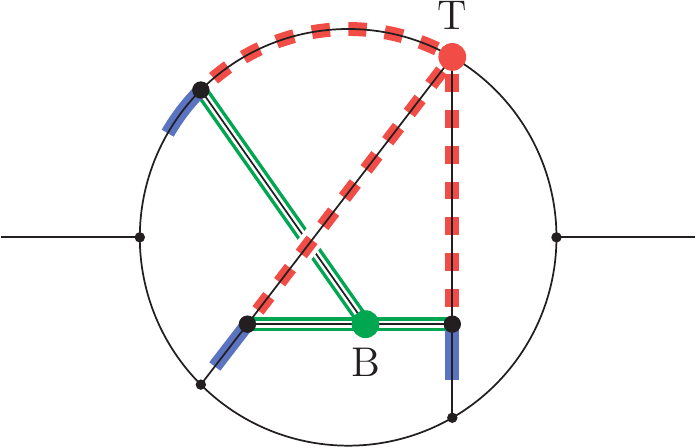}}}
\caption{\label{fig:fiveL}Five-loop topologies}
\end{subfigure}
\caption{\label{fig:diamonds}Two topologies with highlighted diamond 
structures. Red with dashed lines, green with double lines, and blue with 
thick lines represent upper, lower, and external lines of the diamond, 
respectively. Label $T$ represents the top vertex, and $B$ the bottom 
vertex. (a) shows a four-loop topology which can be completely reduced. (b) 
shows the three-loop NO master topology, for which a modified form of the 
diamond rule can be applied to lower the power of line $p_1$ to 1.
(c) shows five-loop topologies, which the diamond rule can be applied to.}
\end{figure}

Several examples of diamond structures are displayed in Fig.~\ref{fig:diamonds}. 
The role of each line in the diamond rule is 
highlighted by different colors and shapes. Red dashed lines, green 
double lines, and blue thick lines represent upper, lower, and 
external lines of the diamond, respectively. Label $T$ represents the top 
vertex, and $B$ the bottom vertex. In Fig.~\ref{fig:no5} a four-loop 
diagram is displayed. For this diagram, the line of either $p_5$, $p_6$, 
$p_7$, $p_8$, $p_9$, or $p_{10}$ can be removed by recursive use of the 
diamond rule or by the explicit formula given in the previous section. The 
irreducible numerators of this diagram are selected as $Q\mydot p_8$, 
$Q\mydot p_{10}$, $p_5 \mydot p_{10}$, and $p_5 \mydot p_7$, such that they 
adhere to the tensorial structure in the diamond rule. The last numerator, 
$p_5 \mydot p_7$, lies outside of the diamond and does not interfere with 
the rule.

If, in this figure, we draw an additional line from the top (T) to the 
bottom (B) vertex, we obtain the simplest nontrivial propagator topology 
with a spectator line. As a five-loop diagram it is unique.

In Fig.~\ref{fig:no} the three-loop master topology NO is displayed. 
$Q \mydot p_5$ is chosen as irreducible numerator. One of the lines attached to the diamond is actually 
an off-shell external line. In general, if the line with momentum $p_{L+1}$
is one of the external momenta of the larger graph, the factor 
$(p_{L+1}^2+m_{L+1}^2)$ is just a constant with respect to the loop 
integration and has no role for reducing the complexity of the integral. 
As a result, the rule~\eqref{eq:diamond-rule} is not applicable to remove one 
of the internal lines. 
Even for such cases, one can still find a useful rule by
shifting $a_{L+1} \to a_{L+1}-1$:
\begin{equation}
  1
  =
  \frac{1}{p_{L+1}^2+m_{L+1}^2} \Biggl[
    \sum_{i=1}^L \frac{a_i}{a_{L+1}-1} \bm{A}_i^+ \bm{A}_{L+1}^-
      ( \bm{B}_i^- - \bm{C}_i^- )
    - \frac{E + 1}{a_{L+1}-1}
      \bm{A}_{L+1}^-
    + \bm{B}_{L+1}^-
  \Biggr]
  ,
  \label{eq:diamond-rule2}
\end{equation}
which decreases at least $a_{L+1}$ or $b_{L+1}$ by one.
Repeated use of this rule from positive integer $a_{L+1}$ and $b_{L+1}$
reduces $a_{L+1}$ or $b_{L+1}$ to $1$. For the NO topology, this variant 
yields the rule to reduce the line $p_1$ to 1 in 
Mincer~\cite{Gorishnii:1989gt,Larin:1991fz}.\footnote{%
The triangle rule counterpart of this variant was used to reduce
the peripheral lines of the massless two-loop propagator-type diagrams with
non-integer powers of the central line to unity.
}%

In Fig.~\ref{fig:fiveL} we show two five-loop topologies for which the 
diamond rule can eliminate one line. The first diagram is unique in the 
sense that it is the simplest diagram for which $L=3$, $S=0$. The second 
diagram is a typical representative of the 29 five-loop topologies with 
$L=2$, $S=0$ and all three $p$-momenta of the diamond internal.

\section{Conclusion and discussion}
\label{sec:conclusion}
We have indicated an extensible, multi-loop topology substructure that can 
be reduced efficiently. We call the corresponding reduction formula the 
diamond rule. Additionally, we have derived an explicit summation formula 
for the recursion in the diamond rule, which avoids spurious poles.

For parametric reduction applications such as the Mincer program, an 
implementation of the diamond rule would be faster than automatically 
generated reduction rules. This is already the case with the summed
triangle as it is used in the Mincer program. It allows the 
program to avoid spurious poles altogether and hence it can run at a fixed 
precision in powers of $\epsilon$.
It is currently not clear whether Laporta approaches~\cite{Laporta:2001dd} 
as used in systems such as AIR~\cite{Anastasiou:2004vj}, 
Reduze~\cite{Studerus:2009ye,vonManteuffel:2012np}, 
and FIRE~\cite{Smirnov:2008iw,Smirnov:2013dia,Smirnov:2014hma} 
benefit from applications of the diamond rule.

It is important to note that the tensorial structure for the irreducible 
numerators should be adhered to. If one chooses numerators of the form 
$(p_i-p_j)^2$ instead of dot products $p_i \mydot p_j$, 
extra terms are introduced to the diamond rule.
It is unclear to us whether the performance gain of using the rule is 
greater than the cost of rewriting the numerators to dot products in software 
such as LiteRed~\cite{Lee:2012cn,Lee:2013mka}. 
In the Mincer approach with its dot products in the 
numerators, the diamond rule fits in perfectly.

Neither at the four-loop level, nor at the five-loop level we have found 
structures that can be reduced by a single IBP identity, apart from the 
diamond rule.

\section*{\label{sec:ac}Acknowledgments}
We like to thank R. Rietkerk, M. Ritzmann, and F. Herzog for discussions.
This work is supported by the ERC Advanced Grant no. 320651, ``HEPGAME''.

\section*{References}

\bibliography{mybibfile}

\end{document}